\documentclass{article}
\usepackage[T1]{fontenc}
\usepackage[utf8]{inputenc}
\usepackage{authblk}

\usepackage{graphicx}
 \usepackage{epsfig}
\usepackage{mathrsfs}
\usepackage{comment}
\usepackage{bm}
\usepackage{epstopdf}
\usepackage{infix-RPN}

\usepackage{amssymb, amsmath, amsthm, complexity,bm}
\usepackage{microtype,dsfont}
\usepackage{tikz,pgfplots,float,subfig,xcolor,cprotect}
\usetikzlibrary{external}                                                                                            
\usepackage{hyperref}
\usepackage{algorithm,algorithmic}

\graphicspath{{}}

\DeclareMathOperator*{\argmin}{\arg\!\min}
\DeclareMathOperator*{\argmax}{\arg\!\max}

\def\R{\mathbb{R}}
\def\NN{\mathbb{N}}

\def\X{\mathcal{X}}
\def\O{\mathcal{O}}
\def\N{\mathcal{N}}

\def\abs#1{\left\lvert#1\right\rvert}
\def\norm#1{\left\lVert#1\right\rVert}

\def\wh#1{\widehat{#1}}
\def\mat#1{\mathbf{#1}}
\def\var#1{{\ttfamily#1}}
\long\def\comment#1{}

\DeclareMathOperator*{\sech}{sech}

\title{Can Small Islands Protect Nearby Coasts From Tsunamis? An Active Experimental Design Approach}

\title{Can Small Islands Protect Nearby Coasts From Tsunamis? An Active Experimental Design Approach}

\author[1,2]{Themistoklis~S.~Stefanakis}%^{1,2}, Emile~Contal^{1},Nicolas~Vayatis^1, Fr\'ed\'eric~Dias^{2,1}, Costas~E.~Synolakis^{3,4}}
\author[1]{Emile~Contal}
\author[1]{Nicolas~Vayatis}
\author[2,1]{Fr\'ed\'eric~Dias}
\author[3,4]{Costas~E.~Synolakis}

\affil[1]{CMLA, ENS Cachan, CNRS, 61 Avenue du Pr\'esident Wilson, F-94230 Cachan, France}
\affil[2]{UCD School of Mathematical Sciences, University College Dublin, Belfield, Dublin 4, Republic of Ireland}
\affil[3]{Hellenic Center for Marine Research, Anavyssos Attikis, GR-19013, Greece}
\affil[4]{Viterbi School of Engineering, University of Southern California, Los Angeles, CA 90089-2531, USA}

\begin{document}
\maketitle  %% Please note that for the copernicus2.cls this command needs to be inserted after \abstract{TEXT}

%\address{$^1$Centre des Math\'ematiques et de Leurs Applications\\
%Ecole Normale Sup\'erieure de Cachan\\
%Cachan, F-94230, France\\
% e-mail: themistoklis.stefanakis@cmla.ens-cachan.fr}
% %
%\address{$^2$School of Mathematical Sciences\\
%University College Dublin\\
%Belfield, Dublin 4, Ireland\\
%e-mail: frederic.dias@ucd.ie; web page: http://mathsci.ucd.ie/~dias/}
%%
%\address{$^3$Hellenic Center for Marine Research\\
%Anavyssos Attikis, GR-19013, Greece}
%%
%\address{$^4$Viterbi School of Engineering\\
%University of Southern California\\
%Los Angeles, CA 90089-2531, USA\\
%e-mail: costas@usc.edu; web page: http://cee.usc.edu/faculty-staff/faculty-directory/synolakis-costas.htm}

\begin{abstract}
Small islands in the vicinity of the mainland are believed to offer protection from wind and waves and thus coastal communities have been developed in these areas. However, what happens when it comes to tsunamis is not clear. Will these islands act as natural barriers ? Recent post-tsunami survey data,  supported by numerical simulations, reveal that the run-up on coastal areas behind small islands was significantly higher than on neighboring locations not affected by the presence of the island. To study the conditions of this run-up amplification, we solve numerically the nonlinear shallow water equations (NSWE). We use the simplified geometry of a conical island sitting on a flat bed in front of a uniform sloping beach. By doing so, the experimental setup is defined by five physical parameters, namely the island slope, the beach slope, the water depth, the distance between the island and the plane beach and the incoming wavelength, while the wave height was kept fixed. The objective is twofold: Find the maximum run-up amplification with the least number of simulations. To achieve this goal, we build an emulator based on Gaussian Processes to guide the selection of the query points in the parameter space. 
\end{abstract}

%--------------------------------------------------------------------------------------------------------------------------------
%                                                              INTRODUCTION
%--------------------------------------------------------------------------------------------------------------------------------

\section{Introduction}  %% \introduction[modified heading if necessary]

In recent years we have witnessed the dreadful damage tsunamis caused in coastal areas around the globe. Especially during the last decade two of the most catastrophic tsunamis ever recorded, the December 2004 tsunami in Indonesia \cite{Liu2005, Titov} and the most recent March 2011 event in Japan \cite{ Fujii2011, Ide2011, Mori2011, Fritz2012} spread panic and pain combined with a huge economic loss at the damaged sites. On the positive side, increased public attention to tsunamis has raised awareness and preparedness, which is the only effective countermeasure and has saved lives, like during the Chilian tsunami in March 2010 \cite{Peachey2010}.

By better understanding the generation, evolution and run-up of tsunami waves, scientists should ultimately provide early warnings and education to coastal communities. Run-up is defined as the maximum wave uprush on a beach or structure above still water level. Since the 1950's tsunami run-up on a plane beach has been extensively studied by \cite{CG58, Keller1964, Synolakis1987, TS94, Brocchini1996, Didenkulova2008, Antuono2010} and numerically by \cite{Stefanakis2011} among others. All these studies deal with the mathematical description of long wave run-up on uniform sloping beaches. The catastrophe in Babi Island \cite{Yeh1993, Yeh1994} focused scientists' attention on tsunami run-up on islands and the studies that followed, which included both laboratory experiments \cite{Briggs1995} and analytical models \cite{Kanoglu1998}, showed that long waves can cause extensive run-up on the lee side of a conical island.
Earlier studies \cite{Homma1950, Longuet1967, Vastano1967, Lautenbacher1970, Smith1975} have given some insight on the behavior of long waves around conical islands, but did not deal with run-up. The big conclusion of all the aforementioned studies  is the fact that long waves do not behave as wind generated waves and that small islands which would act as natural barriers in normal sea conditions, transform into amplifiers of wave energy in areas believed to be protected and where coastal communities thrive. Furthermore, recent findings \cite{Hill2012} have shown enhanced tsunami run-up in areas which lied behind small islands in the vicinity of the mainland and therefore were supposedly protected.

\begin{figure}
\begin{center}
%\scalebox{0.4} % Change this value to rescale the drawing.
%{
%\begin{pspicture}(-15.0,-5.0)(6.0,3.6)
%\psline[linewidth=0.04cm, linecolor = blue](-10.5, 0.0)(-8.5,0.0)
%\psline[linewidth=0.04cm](-15.0, -5.0)(-10.0,-5.0)
%\psline[linewidth=0.04cm, linestyle=dashed](-10.0,-5.0)(-8.0,-5.0)
%\psline[linewidth=0.04cm](-5.0, -5.0)(0.0,-5.0)
%\psline[linewidth=0.04cm](-10.0,-5.0)(-7.5,3.0)
%\psline[linewidth=0.04cm](-7.5,3.0)(-5.0,-5.0)
%\psline[linewidth=0.04cm, linecolor = blue](-6.5,0.0)(3.5,0.0)
%\psline[linewidth=0.04cm](0.0,-5.0)(6.0,3.6)
%\infixtoRPN{+1.5*2.718^(-3*(x+13.5)*(x+13.5))}
%\psplot[linewidth=0.04, linecolor = blue]%    % plots the sinewave
%     {-15.0}{-10.5}{\RPN}
%\usefont{T1}{ptm}{m}{n}
%\rput(2.3,-0.5){\fontsize{30}{30} $\theta_b$}
%\psline[linewidth=0.04cm,arrowsize=0.07cm 2.0,arrowlength=2.4,arrowinset=0.4]{->}(-14.5,2.0)(-12.5,2.0)
%\usefont{T1}{ptm}{m}{n}
%\rput(-9.3,-4.5){\fontsize{30}{30} $\theta_i$}
%\psline[linewidth=0.04cm,arrowsize=0.07cm 2.0,arrowlength=2.4,arrowinset=0.4]{<->}(-11.5,-5.0)(-11.5,0.0)
%\usefont{T1}{ptm}{m}{n}
%\rput(-12.0,-2.5){\fontsize{30}{30} $h$}
%\psline[linewidth=0.04cm, linestyle=dashed](-5.0,-5.0)(-5.0,-4.3)
%\psline[linewidth=0.04cm, linestyle=dashed](0.0,-5.0)(0.0,-4.3)
%\psline[linewidth=0.04cm,arrowsize=0.07cm 2.0,arrowlength=2.4,arrowinset=0.4]{<->}(-5.0,-4.5)(0.0,-4.5)
%\usefont{T1}{ptm}{m}{n}
%\rput(-2.5,-4.0){\fontsize{30}{30} $d$}
%\end{pspicture} 
%}
\includegraphics[height=4.5cm]{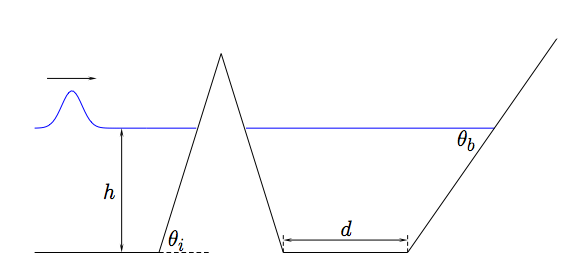}
\end{center}
\caption{Schematic of the geometry of the experimental setup.}
\label{fig:geometry}
\end{figure}

\begin{table}[t]
\caption{Physical parameter ranges}
\vskip4mm
\centering
\begin{tabular}{lcr}
%\tophline
$\tan \theta_i$ & &$0.05 - 0.2$\\
%\middlehline
$\tan \theta_b$ & &$0.05 - 0.2$\\
%\middlehline
d & &$0 - 5000$m\\
%\middlehline
h & &$100 - 1000$m\\
%\middlehline
$\omega$ & &$0.01 - 0.1$rad/s\\
%\bottomhline
\end{tabular}
\label{tab:ranges}
\end{table}

In recent years, the developments in computer science and the increase of computational power in combination with the smaller associated cost compared to laboratory experiments, has led scientists to more and more rely on numerical simulations. However, each simulation has a computational cost, which increases with model complexity and spatiotemporal resolution. Therefore, a series of experiments which have a specific objective, such as maximization/minimization of an output, should be carefully designed in order to reach the desired conclusion with the least number of experiments. Thus, finding the $\argmax_x f(x)$ where $f(x)$ is the output of the experiment depending on the parameters $x$ is not trivial since we do not know the analytical expression of $f(x)$ and therefore it should be approximated. The difficulty of the problem increases with the number of parameters on which the output depends and the ``naive" approach to create regular grids and test all the points becomes prohibitively expensive.  

For this reason, sampling techniques have been developed which aim to reduce the number of points by finding a representative sample of the input space. These techniques are commonly referred to as ``Experimental Design" - e.g. Sacks et al. \cite{Sacks89} - and are static, meaning that the design (sampling) is made initially, before the execution of the experiments and the selection of the future query points is not guided by the experimental results. At the end all the points are queried.  This is already a great advancement compared to the regular grids approach.

More recently, adaptive design \cite{Santner03} and machine (or statistical) learning learning algorithms have been developed for the ``Active Experimental Design", which uses the existing experimental results as a guide for the selection of the future query points, such as Grammacy \& Lee \cite{Gramacy09}. To achieve that, a statistical model $\hat{f}(x)$ of the experiment is built, which is constantly updated as new results arrive. Using the predictions of this statistical model (emulator) the future query points are selected according to the objective of the experiment until we can confidentially say that the objective has been achieved. This dynamic approach can further reduce the computational cost. Moreover, building an emulator has further advantages, the most important one being the ability to use it instead of the actual simulator since it is much less computationally demanding to evaluate and thus can be applied very rapidly, especially in cases where someone needs a quick forecast. Depending on the emulator, it also possible to perform a sensitivity analysis of the model output to the several input parameters. A recent and probably the first example of an emulator built in the context of tsunami research is that of Sarri et al. \cite{Sarri2012}, who emulated landslide-generated tsunamis on a plane beach based on the theoretical model of Sammarco \& Renzi \cite{Sammarco2008}.

From a physical point of view, the current study aims to elucidate the tsunami run-up on a plane beach behind a small conical island compared to an adjacent lateral location on the beach not directly influenced by the presence of the island. To achieve that we use numerical simulations of the nonlinear shallow water equations (NSWE). Moreover, we will present a newly developed method for Active Experimental Design \cite{Contal2013}, which we will apply to our problem and we will also discuss its advantages and limitations in a more general setting. Finally, we will present some metrics that can be used for the comparison of the performance of different learning strategies and an empirical stopping criterion - i.e. a criterion which will signal the achievement of the optimization objective.

%--------------------------------------------------------------------------------------------------------------------------------
%                                                              PROBLEM  SETUP
%--------------------------------------------------------------------------------------------------------------------------------

\section{Experimental Configuration}

\subsection{Simulations}

The simplified bathymetric profile consists of a conical island sitting on a flat bottom and a plane beach behind the island (Fig. \ref{fig:geometry}). The height of the crest of the island above still water level is always fixed at $100$m. The distance between the seaward boundary and the toe of the island is also fixed at $7600$m. A single wave profile is prescribed as forcing at the seaward boundary, having the form $\eta_0(t) = 1.5\sech^2(\omega t - 2.6)$. We use this formulation because we want to avoid the solitary wave link between the water depth and the wave amplitude as is discussed in \cite{Madsen2010}. The problem is governed by $5$ physical parameters, namely the island slope, the plane beach slope, the water depth, the distance between the island and the beach and the prescribed incident wavelength which is controlled by $\omega$ (Table \ref{tab:ranges}).

%% ONE-COLUMN TABLE

%t

\begin{figure}[t]
\vspace*{2mm}
\begin{center}
\includegraphics[height=6.5cm]{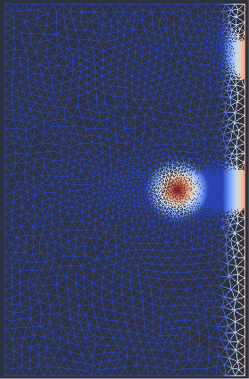}
\end{center}
\caption{The unstructured triangular grid. Colors represent bathymetric contours. The areas of high grid density on the beach, are the locations of run-up measurements.}
\label{fig:mesh}
\end{figure}

The numerical simulations were performed using VOLNA \cite{VOLNA} which solves the NSWE. VOLNA can simulate the whole life cycle of a tsunami from generation to run-up. It uses a Finite Volume Characteristic Flux scheme \cite{Ghidaglia1996, Ghidaglia2001} with a MUSCL type of reconstruction for higher order terms \cite{Kolgan1972, Kolgan1975, Leer1979} and a third order Runge-Kutta time discretization. The code uses an unstructured triangular mesh, which can handle arbitrary bathymetric profiles and can also be refined in areas of interest. The mesh resolution that we used varied from $500$m at the seaward boundary to $2$m at the areas where we measured run-up (Fig. \ref{fig:mesh}).

The run-up was measured on the plane beach exactly behind the island and on a lateral location on the beach, which was far enough from the island and thus was not directly affected by its presence (Fig. \ref{fig:mesh}). To compute run-up, $11$ equally spaced virtual wave gauges were positioned at each location. The gauge location has an inherent uncertainty due to spatial discretization, which is minimized with the use of higher resolution around these locations (Fig. \ref{fig:mesh}). The actual horizontal spacing of the wave gauges was dependent on the beach slope. The minimum height of the gauges was the still water level and the maximum height was selected to be 5.5m above the undisturbed water surface. The run-up never exceeded this height in any of the simulations. The maximum run-up is defined as the maximum recorded wave height at the highest wave gauge. When the wave did not reach the height of a gauge, then that gauge did not record any signal.

\subsection{Experimental Design}
In order to fill the input parameter space we had to choose the input points in such a way that maximal information is obtained with a moderate number of points. This procedure is known as ``Experimental Design" and it is a passive approach as we described in the Introduction. This is the first step to reduce the computational cost. For this purpose we used Latin Hypercube Sampling \cite{McKay1979} with maximization of the minimum distance between points. When using the Latin Hypercube Sampling (LHS) of a function of M variables, the range of each variable is divided into N equally probable, non-overlapping intervals. Then one value from each interval is randomly selected for every variable. Finally, a random combination of N values for M variables is formed. The maximization of the minimum distance between points is added as an extra constraint. The LHS is found to lead to better predictions than regular grids when used with multivariate emulators \cite{Urban2010}. In order to accurately cover the input space, we ran 200 simulations selected by LHS. For comparison a regular grid approach with $10$ grid points in each dimension would require $10^5$ points/simulations. However, we can further improve the performance of  this approach with the so called ``Active Experimental Design", which can suggest the order in which the points will be queried and is result driven, as we describe in the following section. We should clarify that the Active Experimental Design does not require to be initialized with LHS and it can work on any set of points. The LHS was used for evaluation purposes, to better fill the parameter space with a limited number of points, because we do not have the luxury to employ a large number of random points. Finally, we need to stress that in terms of statistical learning, our strategy is not restrained to the specific tsunami research problem and thus can be applied to a wide spectrum of disciplines where the objective is scalar optimization with cost constraints.  

%---------------------------------------------------------------------------------------------------------------------------------------
%                                                                   METHODOLOGY
%---------------------------------------------------------------------------------------------------------------------------------------

\section{Active Experimental Design}
\label{sec:general_strategy}
\subsection{Active Batch Optimization}
Let $f:\X \to \R$ be an unknown function, with $\X$ a finite subset of $\R^d$,
which we can evaluate at any location $x \in \X$ for a fixed cost.
We address the problem of sequentially finding the maximizer of $f$,
\[x^\star=\argmax_{x \in \X} f(x)~,\]
in the lowest possible number of evaluations.
The arbitrary choice of formulating the optimization problem as a maximization
is without loss of generality,
as we can obviously take the opposite of $f$ if the problem is a minimization one.
We consider the special case where $K$ evaluations of $f$ can be acquired with no increase in cost.
For example when $f$ is the result of a numerical experiment
which can be computed in parallel on a machine with $K$ cores,
and the cost to minimize is computation time.
At each iteration $t$, we choose a batch of $K$ points in $\X$ called the queries $\{x_t^k\}_{0\leq k<K}$,
and then observe simultaneously the observations of $f$ at these points, potentially noisy,
\[y_t^k = f(x_t^k) + \epsilon_t^k~,\]
where the $\epsilon_t^k$ is independent Gaussian noise $\N(0,\sigma^2)$. The stochasticity in the measurements is due to the discretization, both spatial and temporal and not due to the accuracy of the evaluation model, since we consider it to be deterministic.

%-----------------------------------------------Objective----------------------------------------------------------------------------------
\subsection{Objective}
Assuming that the number of iterations allowed, hereafter called horizon $T$, is unknown,
a strategy has to be good at any iteration.
Care must be taken to tackle the exploration/exploitation tradeoff,
that is balance learning the function $f$ globally with focusing around the predicted maximum.
We aim to minimize the cumulative regret \cite{bandit_survey},
\[R_T^K = \sum_{t<T} \big( f(x^\star) -\max_{k<K} f(x_t^k) \big)~.\]
The loss $r_t^K$ incurred at iteration $t$ is the simple regret for the batch $\{x_t^k\}_{k<K}$ \cite{pureexplo},
defined as
\[r_t^K = f(x^\star) -\max_{k<K} f(x_t^k)~.\]
A strategy is said to be ``no-regret'', when
\[\frac{R_T^K}{T} \xrightarrow[T \to \infty]{} 0~.\]

%---------------------------------------------Gaussian Processes-------------------------------------------------------------------
\subsection{Gaussian Processes}
In order to analyze the efficiency of a strategy, we have to make some assumptions on $f$.
We want extreme variations of the function to have low probability.
Modeling $f$ as a sample of a Gaussian Process (GP) is a natural way
to formalize the intuition that nearby locations are highly correlated.
It can be seen as a continuous extension of multidimensional Gaussian distributions.
We said that a random process $f$ is Gaussian with mean function $m$
and non-negative definite covariance (or kernel) function $k$, denoted by
\begin{align*}
f &\sim GP(m, k)~,\\
\text{where } m &: \X \to \R\\
\text{and } k &: \X \times \X \to \R^+~,
\end{align*}
when for any finite subset of locations
the values of the random function form a multivariate Gaussian random variable
of mean vector $\bm{\mu}$ and covariance matrix $\mat{C}$
given by $m$ and $k$.
That is, $\forall n<\infty,\ \forall x_1, \dots, x_n \in \X$,
\begin{align*}
  (f(x_1),\dots, f(x_n)) &\sim \N(\bm{\mu}, \mat{C})~,\\
  \text{with } \bm{\mu}[x_i] &= m(x_i)\\
  \text{and } \mat{C}[x_i,x_j] &= k(x_i, x_j)~.
\end{align*}

If we have the prior knowledge that $f$ is drawn from a GP with known kernel function $k$,
then, based on the observations of $f$ after $T$ iterations,
the posterior distribution remains a Gaussian process,
with mean $\wh{\mu}_T$ and variance $\wh{\sigma}_T^2$,
which can be computed via Bayesian inference by \cite{gpml},
\begin{align}
  \label{eq:mu}
  \wh{\mu}_T(x) &= \mat{k}_T(x)^\top \mat{C}_T^{-1}\mat{Y}_T\\
  \label{eq:sigma}
  \text{and }\ \wh{\sigma}^2_T(x) &= k(x,x) - \mat{k}_T(x)^\top \mat{C}_T^{-1} \mat{k}_T(x)~.
\end{align}
\begin{eqnarray}
\mat{X}_T  =  \{x_t^k\}_{t<T,k<K} \quad , \quad \mat{Y}_T  =  [y_t^k]_{x_t^k \in \mat{X}_T} \nonumber% is the vector of noisy observations,
\end{eqnarray}
are the set of queried locations and the vector of noisy observations respectively.
\begin{equation*}
\mat{k}_T(x) = [k(x_t^k, x)]_{x_t^k \in \mat{X}_T}
\end{equation*}
is the vector of covariances between $x$ and the queried points and 

\noindent $\mat{C}_T = \mat{K}_T + \sigma^2 \mat{I}~$ with 

\noindent $\mat{K}_T=[k(x,x')]_{x,x' \in \mat{X}_T}~$ the kernel matrix and $\mat{I}$  stands for the identity matrix.

\begin{figure}[t]
\vspace*{2mm}
  \begin{center}
    \includegraphics[height=5.5cm]{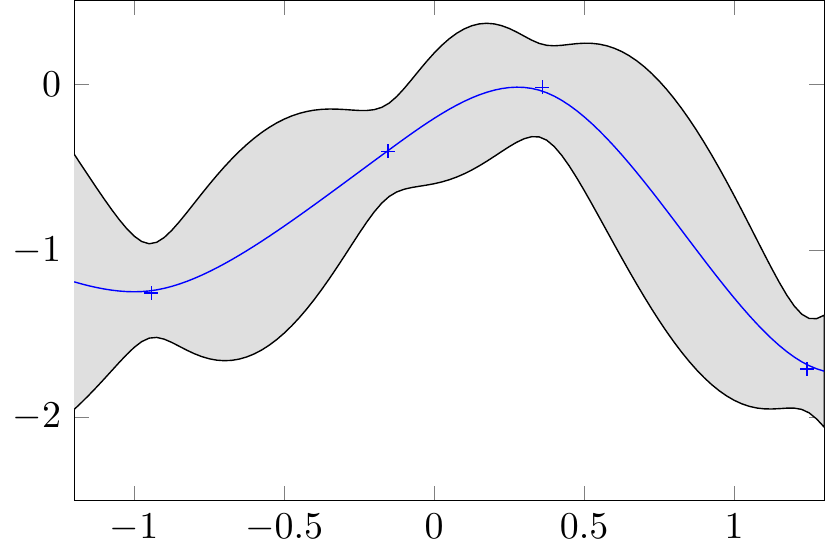}
    \caption{Gaussian Process inference of the posterior mean $\wh{\mu}$ (blue line) and variance $\wh{\sigma}$
      based on 4 realizations (blue crosses).
      The high confidence region (area in grey) is delimited by $\wh{f}^+$ and $\wh{f}^-$.}
    \label{fig:gp}
  \end{center}
\end{figure} 

The three most common kernel functions are:
\begin{itemize}
\item the polynomial kernels of degree $\alpha \in \NN$,
\[\!\!\!\!\!\!\!\!\!\!\!\!k(x_1,x_2) = (x_1^\top x_2 + c)^\alpha~, ~ c \in \R. \]
\item the (Gaussian) Radial Basis Function kernel (RBF or Squared Exponential) with length-scale $l > 0$,
  \begin{align}
    \label{eq:rbf}
    \!\!\!\!\!\!\!\!\!\!\!\!k(x_1,x_2) = \exp\Big(-\frac{\norm{x_1,x_2}^2}{2 l^2}\Big)~,
  \end{align}
\item the Mat\'{e}rn kernel, of length-scale $l$ and parameter $\nu$,
  \begin{align}
    \label{eq:matern} 
    \!\!\!\!\!\!\!\!\!\!\!\!k(x_1,x_2) = \frac{2^{1-\nu}}{\Gamma(\nu)} \left(\frac{\sqrt{2\nu}\norm{x_1,x_2}}{l}\right)^\nu \!\!\!\!K_\nu\Big(\frac{\sqrt{2\nu} \norm{x_1,x_2}}{l}\Big)~,
  \end{align}
  where $K_\nu$ is the modified Bessel function of the second kind and order $\nu$.
 \end{itemize}
The Bayesian inference is represented on Figure \ref{fig:gp} in a sample unidimensional problem.
The posteriors are based on four observations of a Gaussian Process.
The vertical height of the grey area is proportional to the posterior deviation at each point.

%--------------------------------------------------------------------------------------------------------------------------------
%                                           PARALLEL OPTIMIZATION PROCEDURE
%--------------------------------------------------------------------------------------------------------------------------------

\section{Parallel Optimization Procedure}
\label{sec:algo}
Now that we have set the statistical background, we can describe the learning strategy that we used,
namely the Gaussian Process Upper Confidence Bound with Pure Exploration algorithm \cite{Contal2013}.

%-----------------------------------------Confidence Region--------------------------------------------------------------------
\subsection{Confidence Region}
A key property from the GP framework is that the posterior distribution at a location $x$
has a normal distribution $\N(\wh{\mu}_t(x), \wh{\sigma}^2_t(x))$.
We can then define a upper confidence bound $\wh{f}^+$ and a lower confidence bound $\wh{f}^-$,
such that $f$ is included in the interval with high probability:
\begin{align}
  \label{eq:fp}
  \wh{f}_t^+(x) &= \wh{\mu}_t(x) + \sqrt{\beta_t} \wh{\sigma}_{t-1}(x) \quad \text{ and}\\
  \label{eq:fm}
   \wh{f}_t^-(x) &= \wh{\mu}_t(x) - \sqrt{\beta_t} \wh{\sigma}_{t-1}(x)
\end{align}
with $\beta_T \in \O(\log T)$ defined in \cite{Contal2013}. The factor $\beta_t$ regulates the width of the confidence region.

$\wh{f}^+$ and $\wh{f}^-$ are illustrated in Figure \ref{fig:gp} by the upper and lower envelope of the grey area respectively.
The region delimited in that way, the high confidence region,
contains the unknown $f$ with high probability.

\begin{algorithm}[t]
   \caption{\textsf{GP-UCB-PE}}
   \label{alg:algo}
   \begin{algorithmic}
     \STATE $t \gets 0$
     \STATE \var{stop} $\gets \FALSE$
     \WHILE{\var{stop} $ = \FALSE$}
       \STATE Compute $\wh{\mu}_t$ and $\wh{\sigma}_t$ with eq.\ref{eq:mu} and eq.\ref{eq:sigma}
       \STATE \var{stop} $ \gets $ \var{stopping\_criterion}$(\wh{\mu}_{t-1}, \wh{\mu}_t)$
       \STATE $x_t^0 \gets \argmax_{x \in \X} \wh{f}_t^+(x)$
       \STATE Compute $\mathfrak{R}_t$ with eq.\ref{eq:r}
       \FOR{$k=1,\dots,K-1$}
         \STATE Compute $\wh{\sigma}_t^{(k)}$ with eq.\ref{eq:sigma}
         \STATE $x_t^k \gets \argmax_{x \in \mathfrak{R}_t} \wh{\sigma}_t^{(k)}(x)$
       \ENDFOR
       \STATE Query $\{x_t^k\}_{k<K}$
       \STATE $t\gets t+1$
     \ENDWHILE
   \end{algorithmic}
\end{algorithm}

%---------------------------------------Relevant Region--------------------------------------------------------------------------

\subsection{Relevant Region}
We define the relevant region $\mathfrak{R}_t$ being the region
which contains $x^\star$ with high probability.
Let $y_t^\bullet$ be our lower confidence bound on the maximum,
\begin{align*}
  x_t^\bullet &= \argmax_{x \in \X} \wh{f}_t^-(x)\\
  \text{and } y_t^\bullet &= \wh{f}_t^-(x_t^\bullet)~.
\end{align*}
$y_t^\bullet$ is represented by the horizontal dotted green line on Figure \ref{fig:algo}.
$\mathfrak{R}_t$ is defined as,
\begin{align}
\label{eq:r}
\mathfrak{R}_t = \Big\{ x \in \X \mid \wh{f}_t^+(x) \geq y_t^\bullet \Big\}~.
\end{align}
$\mathfrak{R}_t$ discards the locations where $x^\star$ does not belong with high probability.
It is represented in green in Figure \ref{fig:algo}.

%In the sequel, we will use a modified version of the relevant region
%which also contains $\argmax_{x \in \X} \wh{f}_{t+1}^+(x)$ with high probability.
%The novel relevant region is formally defined by
%\begin{align}
%\label{eq:r}
%\mathfrak{R}_t^+ = \Big\{ x \in \X \mid \wh{\mu}_t(x) + 2\sqrt{\beta_{t+1}} \wh{\sigma}_{t-1}(x) \geq y_t^\bullet \Big\}~.
%\end{align}
%Using $\mathfrak{R}_t^+$ instead of $\mathfrak{R}_t$ guarantees that
%the queries at iteration $t$ will leave an impact
%on the future choices at iteration $t+1$.

%-----------------------------------------GP-UCB-PE-----------------------------------------------------------------------------

\subsection{\textsf{GP-UCB-PE}}

\begin{figure}[t]
  \begin{center}
    \includegraphics[height=5.5cm]{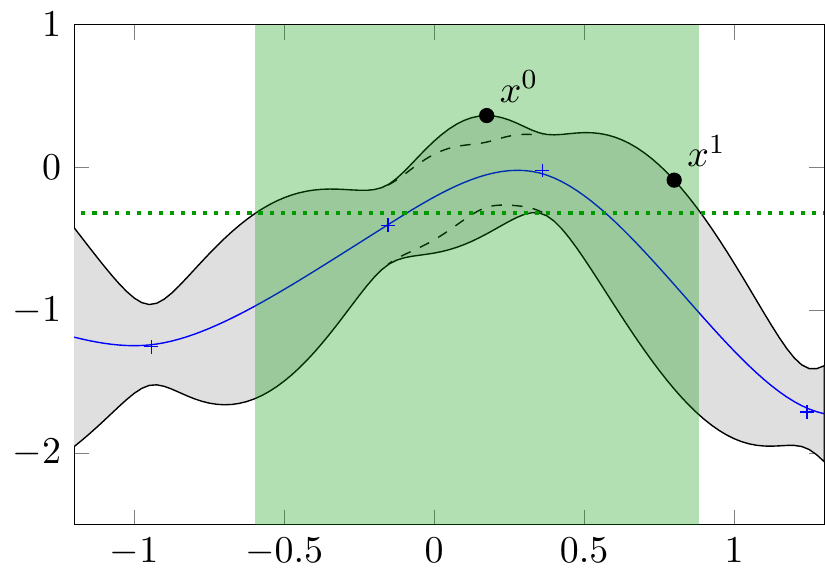}
    \caption{
      Two queries of \textsf{GP-UCB-PE} on the previous example.
      The lower confidence bound on the maximum is represented by the horizontal dotted green line at $y_t^\bullet$.
      The relevant region $\mathfrak{R}$ is shown in light green (without edges).
      The first query $x^0$ is the maximizer of $\wh{f}^+$.
      We show in dashed line the upper and lower bounds with the update of $\wh{\sigma}$
      after having selected $x^0$.
      The second query $x^1$ is the one maximizing the uncertainty inside $\mathfrak{R}$.}
    \label{fig:algo}
  \end{center}
  \vskip -0.2in
\end{figure}

We present here the Gaussian Process Upper Confidence Bound with Pure Exploration algorithm
(\textsf{GP-UCB-PE}), a very recent algorithm from \cite{Contal2013}
combining two strategies to determine the queries $\{x_t^k\}_{k<K}$
for batches of size $K$.
The first location is chosen according to the \textsf{UCB} rule,
\begin{align}
\label{eq:argmax_f}
x_t^0 = \argmax_{x \in \X} \wh{f}_t^+(x) ~.
\end{align}
This single rule is enough to deal with the exploration/exploitation tradeoff.
The value of $\beta_t$ balances between exploring uncertain regions
(high posterior variance $\wh{\sigma}_t^2(x)$)
and focusing on the supposed location of the maximum
(high posterior mean $\wh{\mu}_t(x)$).
This policy is illustrated with the point $x^0$ in Figure \ref{fig:algo}.

The $K-1$ remaining locations are selected via Pure Exploration
restricted to the region $\mathfrak{R}_t$.
We aim to maximize $I_t$, the information gain about $f$ granted by the $K-1$ points \cite{info_theory}. This can be efficiently approximated by the greedy procedure which selects the $K-1$ points one by one
and never backtracks 
\footnote[1]{Formally, $I_t$ is the gain in Shannon entropy $H$
when knowing the values of the observations at those points,
conditioned on the observations we have seen so far, $I_t(\mat{X}) = H(\mat{Y}) - H(\mat{Y} \mid \mat{X}_t)~.$
Finding the $K-1$ points that maximize $I_t$ is known to be intractable \cite{max_entropy} and thus an approximation is required.
}.

For a Gaussian distribution we have $I_t(\mat{X}) \in \O(\log \det \mat{\Sigma})$,
where $\mat{\Sigma}$ is the covariance matrix of $\mat{X}$.
The location of the single point that maximizes the information gain
is easily computed by maximizing the posterior variance.
Our greedy strategy selects for each $1\leq k < K$ the following points one by one,
\begin{align}
\label{eq:argmax_s}
x_t^k = \argmax_{x \in \mathfrak{R}_t} \wh{\sigma}_t^{(k)}(x)  ~,
\end{align}
where $\wh{\sigma}_t^{(k)}$ is the updated variance after choosing $\{x_t^{k'}\}_{k'<k}$.
We use here the fact that the posterior variance does not depend on the values $y_t^k$ of the observations,
but only on their position $x_t^k$.
One such point is illustrated with $x^1$ in Figure \ref{fig:algo}.

These $K-1$ locations reduce the uncertainty about $f$,
improving the guesses of the \textsf{UCB} procedure by $x_t^0$.
The overall procedure is shown in Algorithm \ref{alg:algo}.

%---------------------------------------Regret Bounds----------------------------------------------------------------

\subsection{Theoretical Guarantees}
\label{sec:regret}
With the Gaussian process assumption on $f$,
we can adjust the parameter $\beta_t$ such that $f$
will be contained by the high confidence region with high probability.
Under this condition,
\cite{Contal2013} prove a general bound on the regret achieved by \textsf{GP-UCB-PE},
making this strategy a ``no-regret'' algorithm.
The order of magnitude of the cumulative regret $R_T^K$ we obtained with a linear kernel
is $\sqrt{\frac{T}{K}\log TK}$,
and for RBF Kernel $\sqrt{\frac{T}{K}(\log TK)^d}$, up to polylog factors.
When $K \ll T$, these probabilistic bounds with parallel queries are better
than the ones incurred by the sequential procedure by an order of $\sqrt{K}$.

%---------------------------------------Stopping Criterion-----------------------------------------------------------

\subsection{Stopping criterion}
One challenging problem we face in practice
is to decide when to stop the iterative strategy.
The theoretical analysis only gives general expected bounds when $T$ tends to infinity,
it does not provide estimations of the constant and short term factors.
We have to design an empirical, yet robust, criterion.
One trivial solution is to fix the number of iterations (or computation time) allowed by a predefined limit.
This is not a suitable solution for the general case,
as one does not know precisely the amount of exploration needed
to be confident about the maximum of $f$.
Other rules like the criteria based on the local changes of the queries
in the target space (improvement-based criteria) as well as in the input space (movement-based criteria),
come from the convex optimization literature.
These are not suitable in our global, nonconvex setting,
as they stops after a local maximum is found.

\begin{figure}[t]
  \begingroup
  \tikzset{every picture/.style={scale=.7}}
  \centering
  \includegraphics[height=6.5cm]{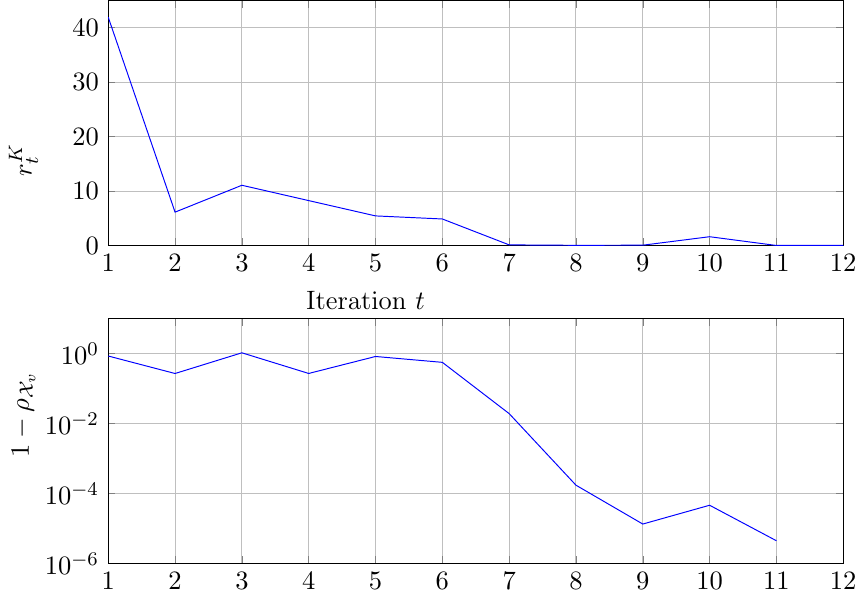}
  \cprotect\caption{Relationship between the simple regret $r_t^K$, unknown in a real situation,
    and the rank correlation $\rho_{\X_v}(\pi_{t-1},\pi_t)$ (in log-scale),
    for the synthetic function \verb/Himmelblau/.
    The stopping threshold $\rho_0$ was set to $10^{-4}$, and the lag $\ell$ to $4$,
    the algorithm stopped at iteration $12$, after having found a good candidate at iteration $7$
    and the true maximum at iteration $11$.
  }
  \label{fig:regret_rank}
  \endgroup
\end{figure}

Our approach is to stop when the procedure ceases to learn relevant information on $f$.
We attempt to measure the global changes in the estimator $\wh{\mu}_t$
between two successive iterations,
with more focus on the highest values.
The algorithm then stops when these changes become insignificant
for a short period.
The change between $\wh{\mu}_t$ and $\wh{\mu}_{t+1}$ is measured
by the correlation between their respective values
on a finite validation data set $\X_v \subset \X$.

We denote by $n_v$ the size of the validation data set $\abs{\X_v}$
and $\mathfrak{S}_{n_v}$ the set of all permutations of $[1 \dots n_v]$.
Let $\pi_t \in \mathfrak{S}_{n_v}$ (resp. $\pi_{t+1}$) be the ranking function
associated to $\wh{\mu}_t$ (resp. $\wh{\mu}_{t+1}$),
such that 
\begin{eqnarray}
\pi_t(\argmax_{x \in \X_v} \wh{\mu}_t(x)) &=& 1~ \quad \text{and}\nonumber \\
\pi_t(\argmin_{x \in \X_v} \wh{\mu}_t(x)) &=& n_v~. \nonumber
\end{eqnarray}
We then define the discounted rank dissimilarity $d_{\X_v}$ and the normalized rank correlation $\rho_{\X_v}$ as
\begin{align*}
d_{\X_v}(\pi_t,\pi_{t+1}) &= \sum_{x \in \X_v} \frac{\big(\pi_{t+1}(x)-\pi_t(x)\big)^2}{\big(\pi_{t+1}(x)\big)^2}~,\\
\rho_{\X_v}(\pi_t,\pi_{t+1}) &=
    1 - \frac{d_{\X_v}(\pi_t,\pi_{t+1})}
             {\max_{\pi^+,\pi^- \in \mathfrak{S}_{n_v}} d_{X_v}(\pi^+,\pi^-)}~.
\end{align*}
The denominator in the definition of $\rho_{\X_v}$ represents the discounted rank dissimilarity
between two reversed ranks $\pi^+$ and $\pi^-$.
The normalized rank correlation for such two ranks will therefore be equal to $0$,
whereas for any rank $\pi$, this correlation with itself will be $\rho_{\X_v}(\pi,\pi)=1$.
The correlation $\rho_{\X_v}$ can be seen as a modified Spearman's rank correlation coefficient,
where the squared distances are weighted by their position in the new rank $\pi_{t+1}$.
If we note $\pi\{n \leftrightarrow m\}$ the inversion of the $n^\text{th}$ and $m^\text{th}$ ranks in $\pi$,
for all rank $\pi$,
we remark that $\rho_{\X_v}(\pi,\pi\{2 \leftrightarrow 3\}) > \rho_{\X_v}(\pi,\pi\{1 \leftrightarrow 2\})$.

We show in Figure \ref{fig:regret_rank} the observed relationship 
between the regret incurred at iteration $t$ and the rank correlation $\rho_{\X_v}(\pi_{t-1},\pi_t)$.
The algorithm stops when $\rho_{\X_v}(\pi_t,\pi_{t+1})$ stays below a given threshold $\rho_0$
for $\ell$ iterations in a row.
The value of this threshold has to be fixed empirically.
In Figure \ref{fig:regret_rank}, $\rho_0=10^{-4}$ and $\ell=4$,
the algorithm stopped at iteration $12$.

\begin{figure*}
  \centering
  \subfloat[][$\rho_0=10^{-2}$]{
    \includegraphics[width=3in]{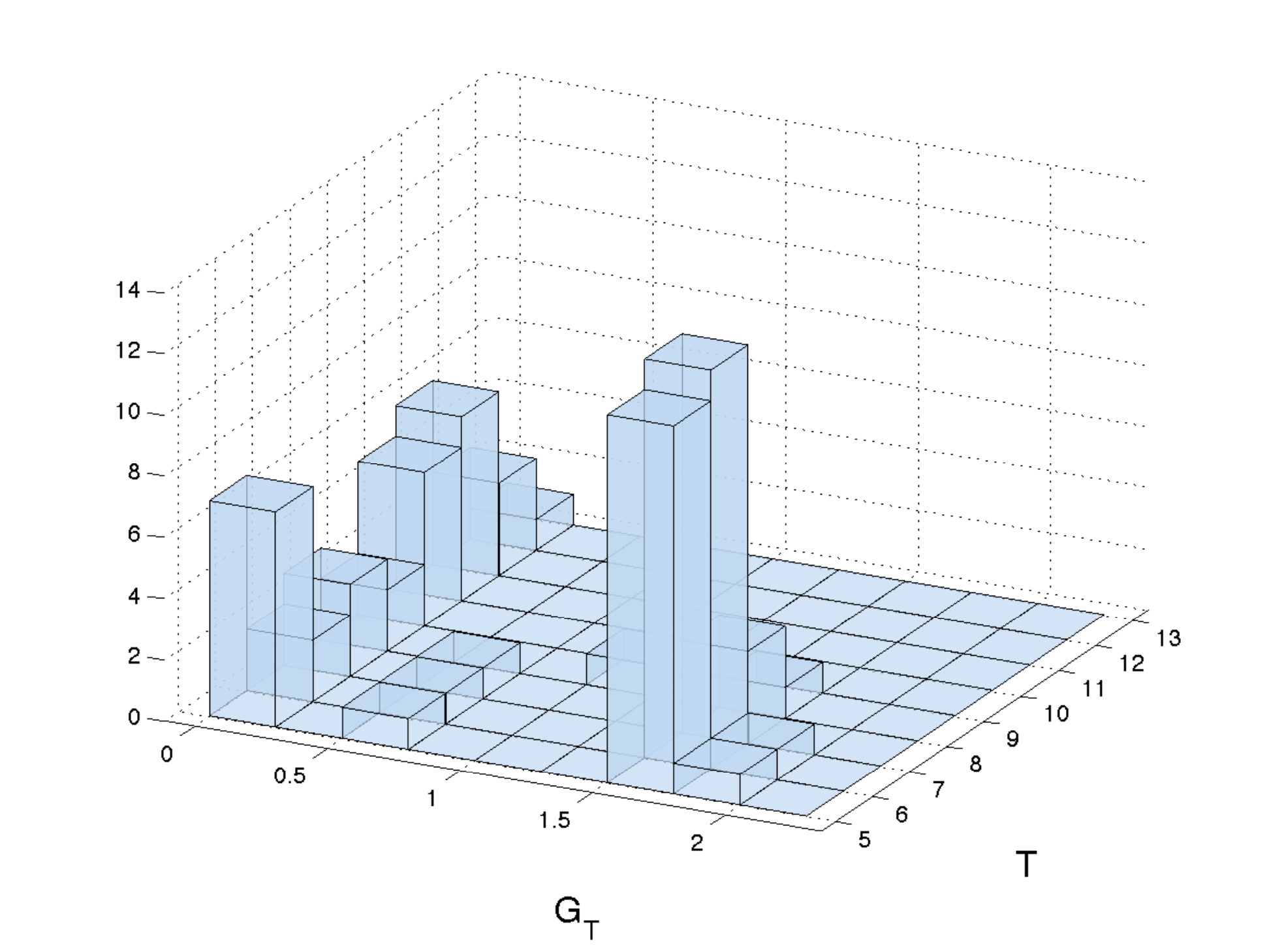}
  }
  \subfloat[][$\rho_0=10^{-3}$]{
    \includegraphics[width=3in]{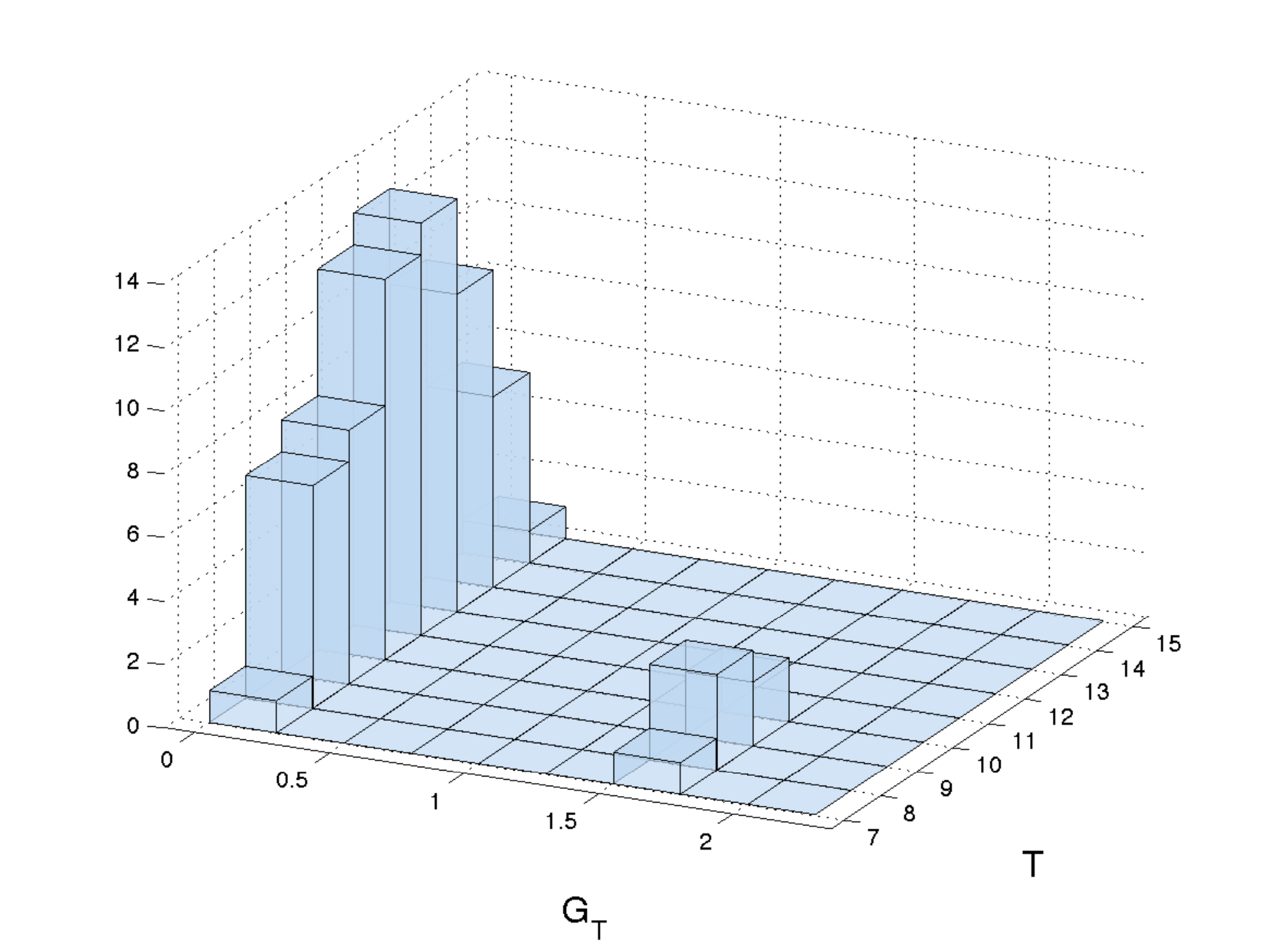}
  }\\
  \subfloat[][$\rho_0=10^{-4}$]{
    \includegraphics[width=3in]{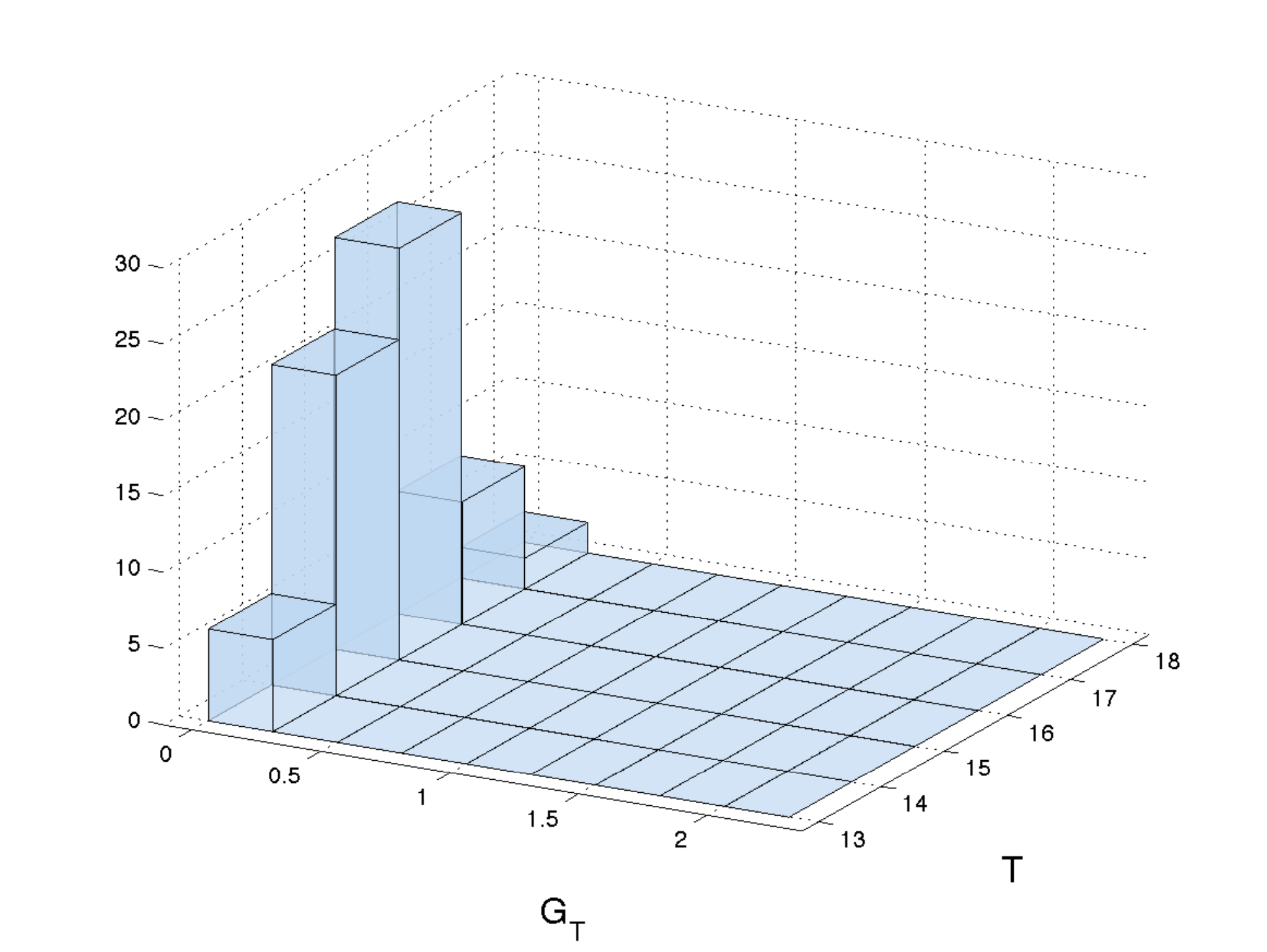}
  }
  \subfloat[][$\rho_0=10^{-5}$]{
    \includegraphics[width=3in]{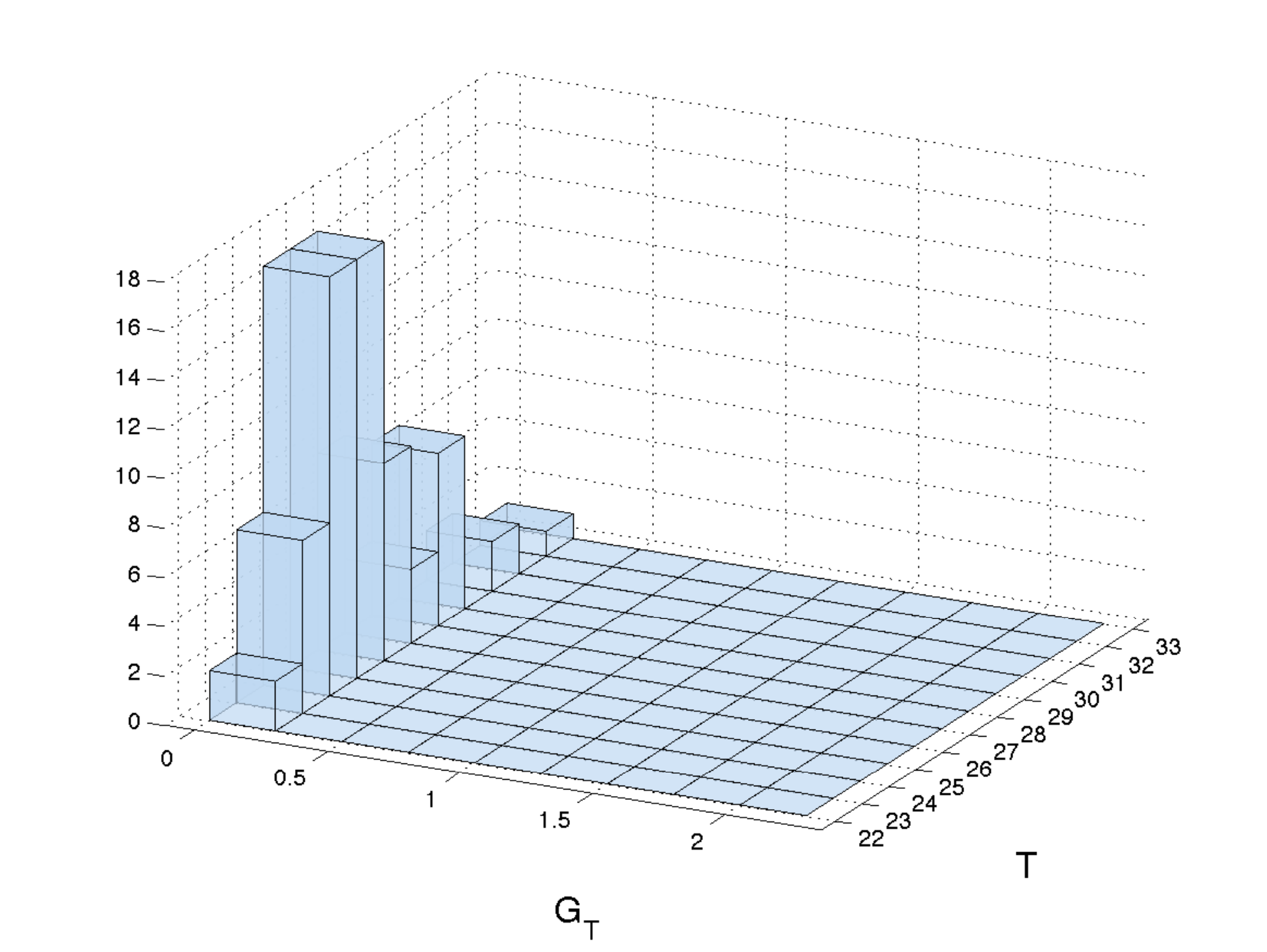}
  }

  \caption{Distribution of the final number of iterations $T$ and the final gap (minimum regret) $G_T$
    for $4$ different thresholds.}
  \label{fig:regret_iteration}
\end{figure*}

In Figure \ref{fig:regret_iteration},
we can see the distribution of the final number of iterations $T$ (lower is better)
together with the final gap $G_T= \min_{t<T} R_t^K$ (lower is better)
for $4$ different thresholds $\rho_0$. The value
$\rho_0=10^{-4}$ appears to be a good threshold,
since the final regret is always $0$, and the number of iterations remains low. Further reduction of this threshold will again guarantee zero regret, but with higher computational cost. On the other hand, greater values of $\rho_0$ will reduce the computational cost, but with an increased risk to miss the maximum.

%--------------------------------------------------------------------------------------------------------------------------------
%                                                              EXPERIMENTS
%--------------------------------------------------------------------------------------------------------------------------------

\section{Experiments}
\label{sec:expes} 

\subsection{Synthetic data sets}

\begin{figure*}[t]
  \begingroup
  %\tikzset{every picture/.style={scale=.7}}
  \centering
  \subfloat[][Himmelblau]{
    \includegraphics[width=3in]{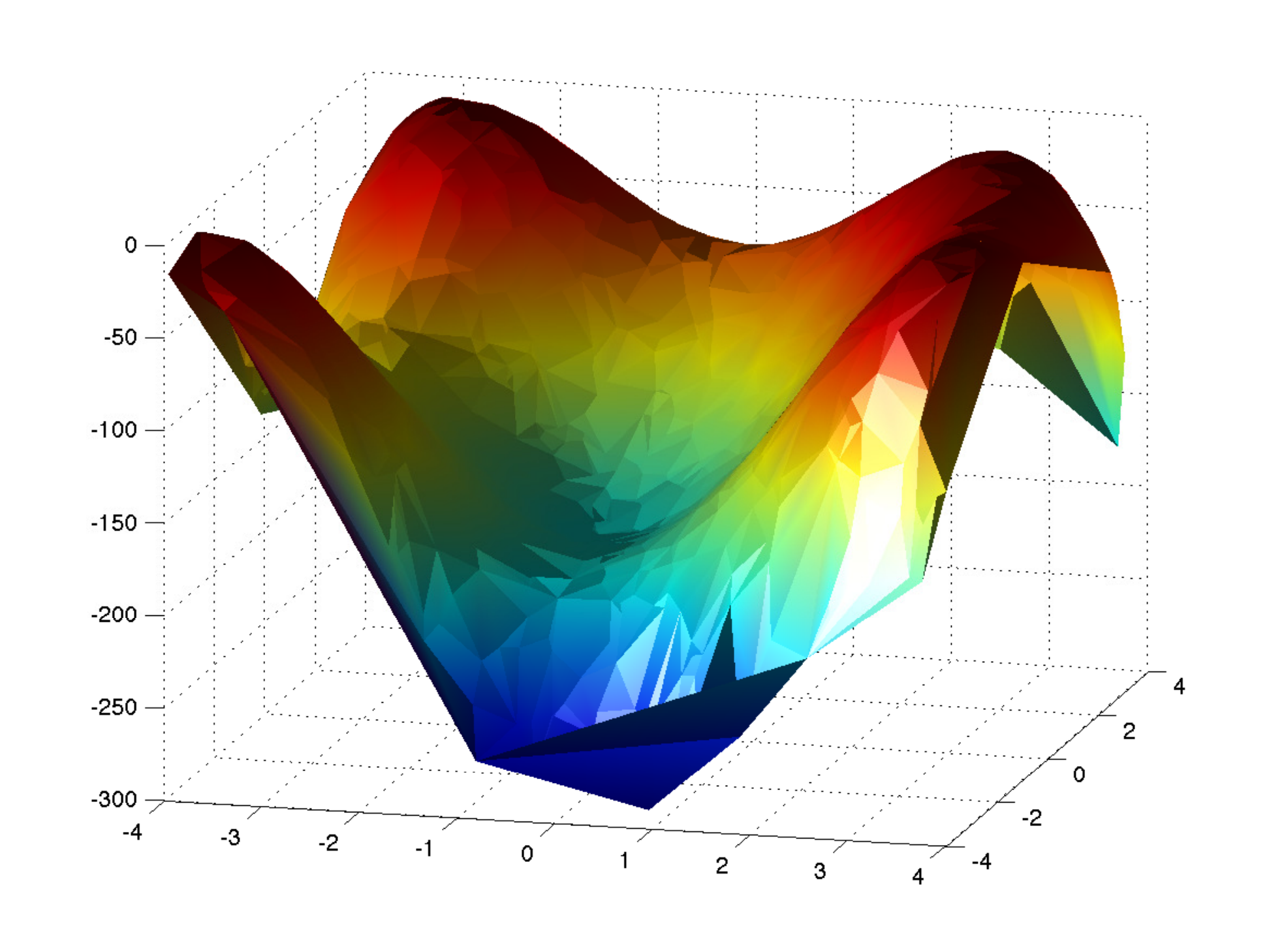}
    \label{fig:himmelblau_3d}
  }
  \subfloat[][Gaussian mixture]{
    \includegraphics[width=3in]{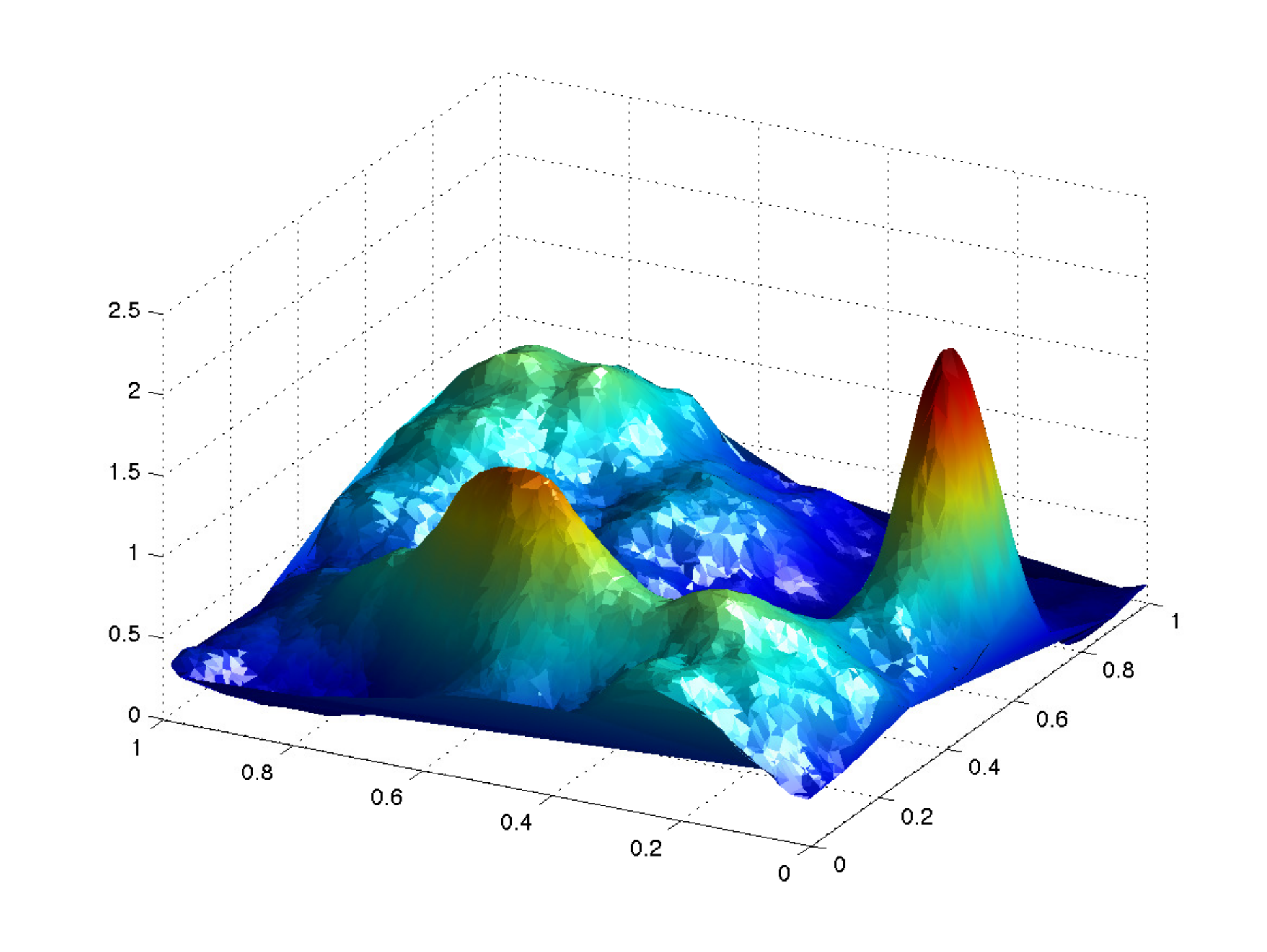}
    \label{fig:random_gaussian_3d}
  }
  \endgroup
  \caption{Visualization of the synthetic data sets used for assessment.}
\end{figure*} 

\begin{figure*}[t]
  %\begingroup
  %\tikzset{every picture/.style={scale=.62}}
  \centering
  \subfloat[][Tsunamis]{
    \includegraphics[width=1.8in]{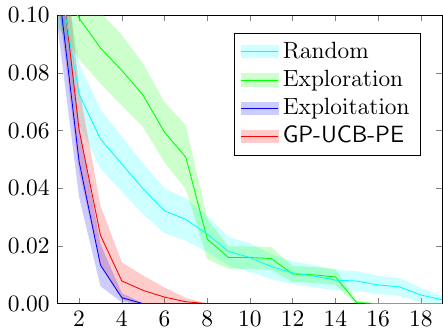}
    \label{fig:expe_tsunamis}
  }
  \subfloat[][Himmelblau]{
    \includegraphics[width=1.8in]{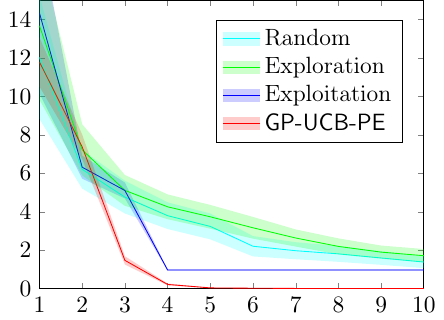}
    \label{fig:expe_himmelblau}
  }
  \subfloat[][Gaussian mixture]{
    \includegraphics[width=1.85in]{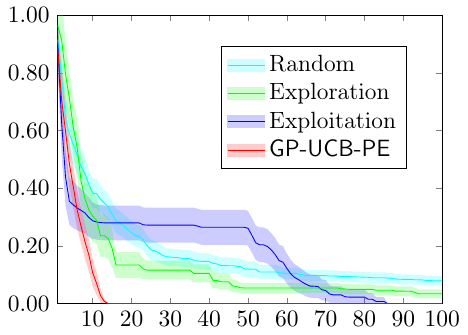}
    \label{fig:expe_gaussian}
  }
  %\endgroup
  \caption{Experiments on several real and synthetic tasks.
    The curves show the decay of the mean of the simple regret $r_t^K$ with respect to the iteration $t$,
    over $64$ experiments.
    We show with the translucent area the 95\% confidence intervals.
    }
  \label{fig:expes}
\end{figure*}

Apart from the tsunami experiment, which is $5$-dimensional and we do not know a priori the form of the response surface, in order to test the performance of the active learning algorithm, it is wise to use some synthetic data sets. These data sets  can be easily visualized ($2$-dimensional) and we can attribute to them some desired properties, such as several local maxima or background noise, which aim to test the algorithm and will give us a direct feedback of its behavior.  
%\paragraph{Generated GP.}
%The \verb/Generated GP/ functions are random GPs drawn from a Mat\'{e}rn kernel
%in dimension $2$, with the kernel bandwidth set to $\frac{1}{4}$,
%the Mat\'{e}rn parameter $\nu=3$ and noise variance $\sigma^2$ set to $1$.

\subsubsection{Himmelblau function}
The Himmelblau data set is a nonconvex function in dimension $2$.
We compute a slightly tilted version of the Himmelblau function,
and take the opposite to match the challenge of finding its maximum.
This function presents four peaks but only one global maximum (near $(-3.8,-3.3)$).
It gives a practical way to test the ability of a strategy to manage exploration/exploitation tradeoffs.
It is represented in Figure \ref{fig:himmelblau_3d}.

\subsubsection{Gaussian mixture}
This synthetic function comes from the addition of three $2$-D Gaussian functions.
at $(0.2, 0.5)$, $(0.9, 0.9)$, and the maximum at $(0.6, 0.1)$.
We then perturb these Gaussian functions with smooth variations
generated from a Gaussian Process with Mat\'{e}rn Kernel (Eq. \ref{eq:matern})
and very few noise.
It is shown on Figure \ref{fig:random_gaussian_3d}.
The highest peak being thin, the sequential search for the maximum of this function
is quite challenging.

%------------------------------------------Assessment--------------------------------------------------------------------

\begin{figure}[t]
  \begingroup
  \tikzset{every picture/.style={scale=.8}}
  \centering
  \includegraphics[height=4.5cm]{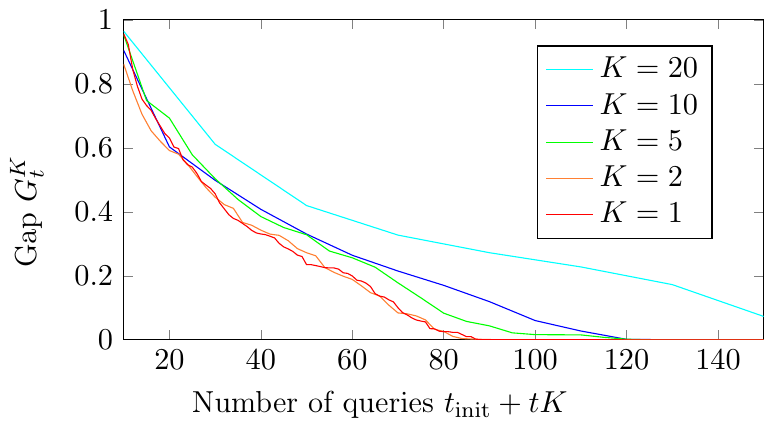}
  \cprotect\caption{Impact of the batch size $K$ on the gap $G_t^K$.                                                  
The curves show the mean of $64$ experiments on the synthetic data \verb/Gaussian Mixture/,                       
with $t_{\text{init}}=10$.
  }
  \label{fig:batch_size}
  \endgroup
\end{figure}

\subsection{Assessment}
We verify empirically the performance of \textsf{GP-UCB-PE}
by measuring the decay of the regret obtained on several synthetic functions.
For the sake of convenience
we do not report the cumulative regret $R_t^K$ on the figures,
but the gap between the maximum point discovered and the true maximum,
defined as the minimum regret so far $G_t^K = \min_{t' \leq t} r_t^k$.

First, we show in Figure \ref{fig:batch_size} the impact of the size of the batch $K$ on the minimum regret $G_t^K$. It is shown that the sequential (the red curve $K = 1$) performs better than the rest, without however being extremely outperforming (orders of magnitude). Therefore in a situation where the final number of queries is the restrictive factor, one would better choose small batch sizes. On the other hand, if total time of the optimization is the restrictive factor, then one could choose larger batch sizes without sacrificing too much computational cost.

We then compare our approach to three other strategies,
\begin{itemize}
\item \verb/Random/, which chooses the next queries $\{x_t^k\}_{k < K}$ at random,
\item \verb/Exploration/, which attempts to maximize the information gain on $f$ at each iteration,
\[x_t^k = \argmax_{x \in \X} \wh{\sigma}_t^{(k)}(x)~,\]
\item \verb/Exploitation/, which only focuses on the predicted maximum,
\[x_t^k = \argmax_{x\in \X \setminus \{x_t^{k'}\}_{k'<k}} \wh{\mu}_t(x)~.\]
\end{itemize}

%--------------------------------------------Protocol--------------------------------------------------------------------

%\subsection{Protocol}
For all data sets and algorithms, the batch size $K$ was set to $10$
and the learners were initialized with a random subset of $t_{\text{init}}=20$ observations $(x_i, y_i)$.
The curves in Figure \ref{fig:expes} show the evolution of the gap $G_t^K$
in term of iteration $t$.
We report the average value with the confidence interval over $64$ experiments (random initializations).
The kernel function used was always an RBF kernel (Eq. \ref{eq:rbf}).
The parameters of the algorithm, like the length-scale of $k$ (represented by $l$),
were chosen as the best parameters found by validation on a random subsample of the data.

Our learning algorithm is shown to outperform the rest strategies on the synthetic data sets. \verb/Exploitation/ in these data sets looses time because it gets stuck in a local maximum, while \verb/Exploration/ and \verb/Random/ strategies will asymptotically find the global maximum on average.  On the \verb/Tsunami/ data set, \verb/Exploitation/ perform slightly better than \textsf{GP-UCB-PE} probably due to the simplicity of the run-up function, which even though is $5$-dimensional does not seem to pose any serious challenges, probably because it has only one maximum. 

\comment{The algorithm \textsf{SM} ---Simulation Matching--- described in \cite{sm},
with \textsf{UCB} base policy,
has shown similar results to \textsf{GP-UCB-PE} on synthetic functions
(Figures \ref{fig:expe_gp}, \ref{fig:expe_himmelblau})
and even better results on chaotic problem without noise (Figure \ref{fig:expe_mg}),
but performs worse on real noisy data (Figures \ref{fig:expe_tsunamis}, \ref{fig:expe_abalone}).
On the contrary, the initialization phase of \textsf{GP-BUCB}
leads to good regret on difficult real tasks (Figure \ref{fig:expe_tsunamis}),
but looses time on synthetic Gaussian or polynomial ones (Figures \ref{fig:expe_gp}, \ref{fig:expe_himmelblau}).
The number of dimensions of the \verb/Abalone/ task
is already a limitation for \textsf{GP-BUCB} with the RBF kernel,
which losses many iterations in the initialization phase.
The mean regret for \textsf{GP-BUCB} converges to zero abruptly after the initialization phase at iteration $55$,
and is therefore not visible on Figure \ref{fig:expe_abalone}.

\textsf{GP-UCB-PE} achieves good performances on both sides.
We obtained better regret on synthetic data
as well as on real problems from the domains of physics and biology.
Moreover, the computation time of \textsf{SM} was two order of magnitude longer than the others.
}

\section{The Effect Of The Conical Island}

\begin{figure}[t]
\vspace*{2mm}
\begin{center}
\includegraphics[width=8.3cm]{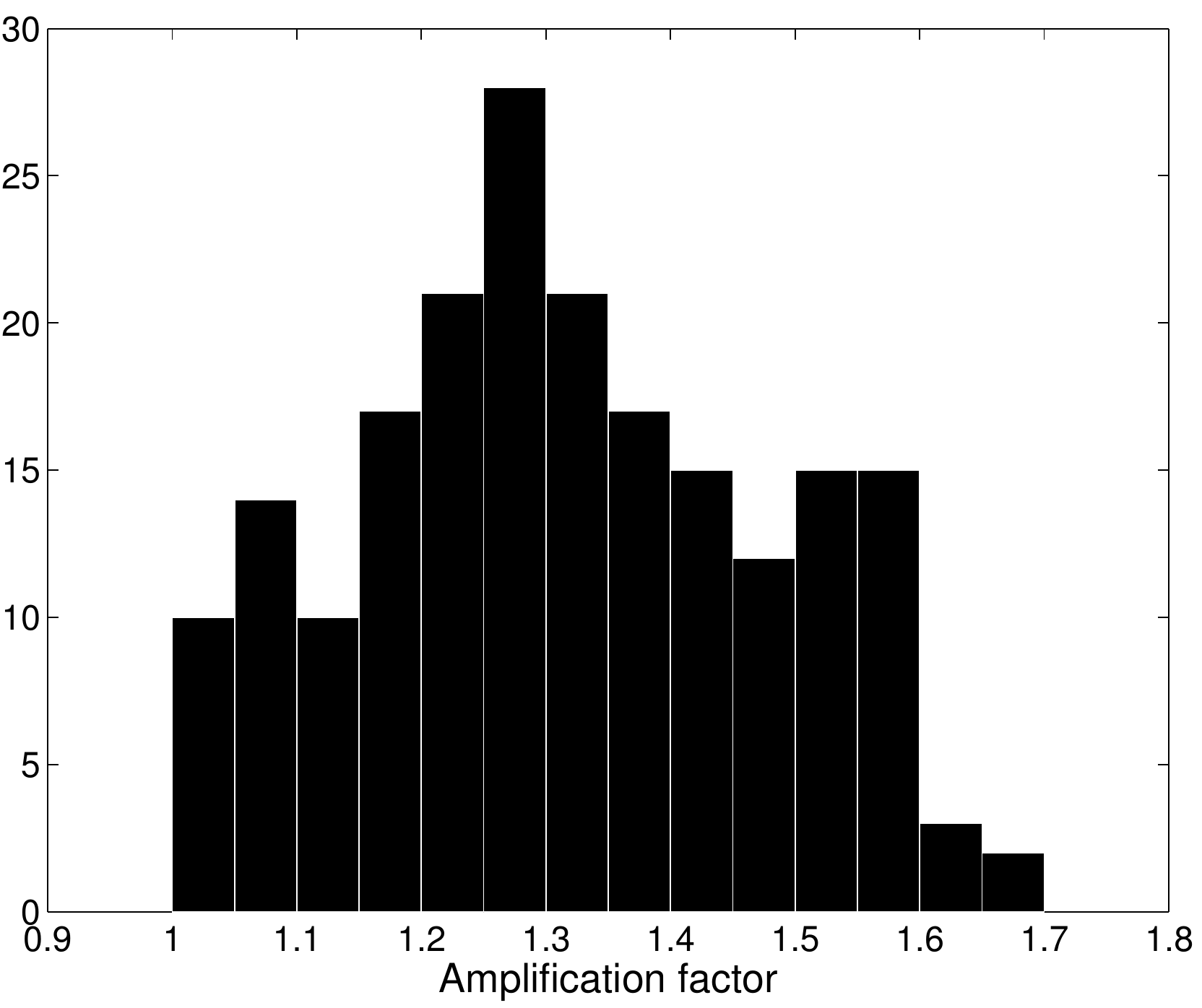}
\end{center}
\caption{Histogram of the run-up amplification on the beach directly behind the island compared to the run-up on a lateral location on the beach, not directly affected by the presence of the island.}
\label{fig:hist}
\end{figure}

%\begin{figure}[t]
%\vspace*{2mm}
%\begin{center}
%\includegraphics[width=8.3cm]{conical_island.png}
%\end{center}
%\caption{View of the wave as its passes the island and heads towards the beach. The island delays the wave in its proximity, while far from it the wave propagates unaffected. The vertical dimension has been magnified for visualization purposes.}
%\label{fig:conical_island}
%\end{figure}

\begin{figure*}
  \centering
  \subfloat[][]{
    \includegraphics[width=0.45\columnwidth]{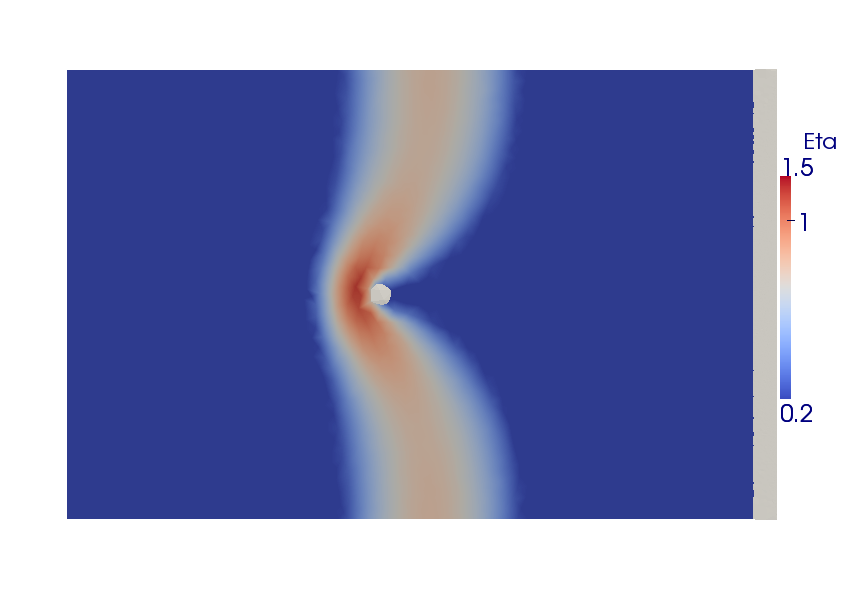}
  }
  \subfloat[][]{
    \includegraphics[width=0.45\columnwidth]{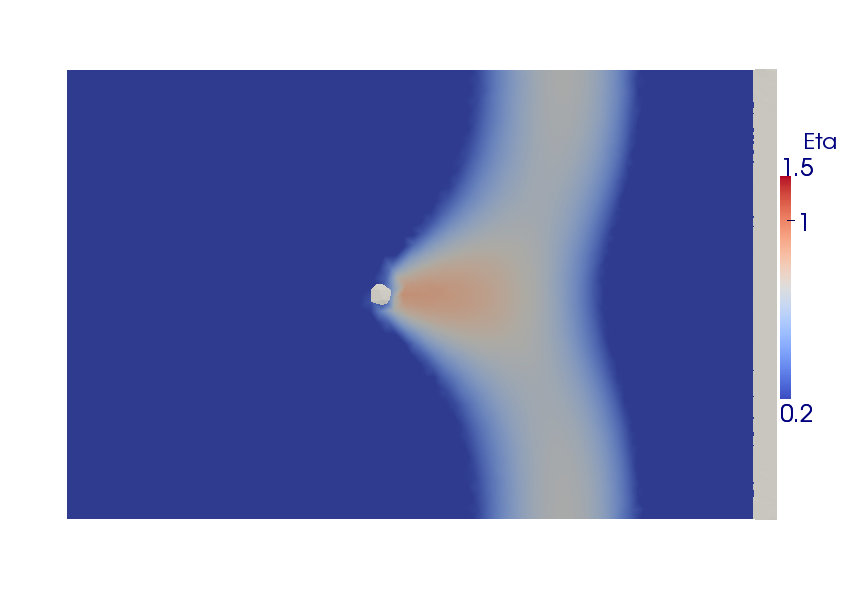}
  }\\
    \subfloat[][]{
    \includegraphics[width=0.45\columnwidth]{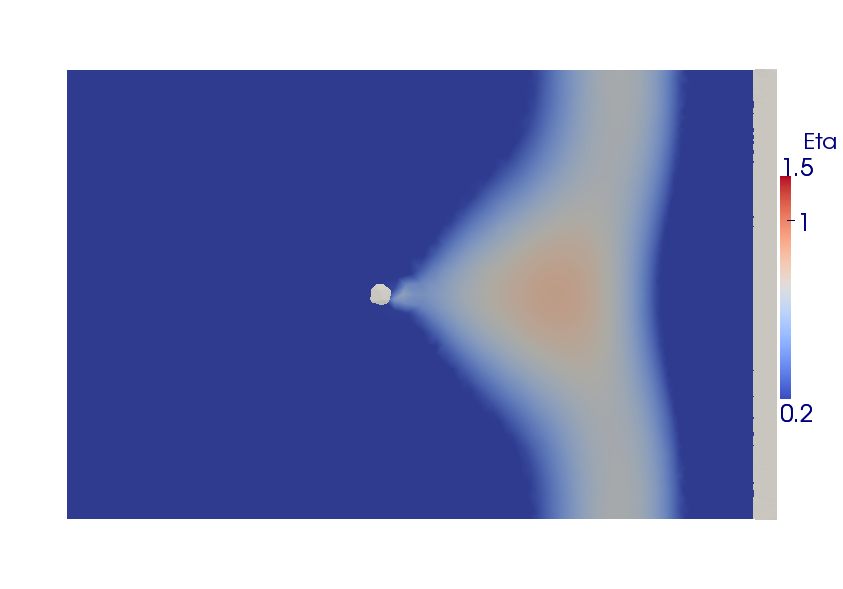}
  }
  \subfloat[][]{
    \includegraphics[width=0.45\columnwidth]{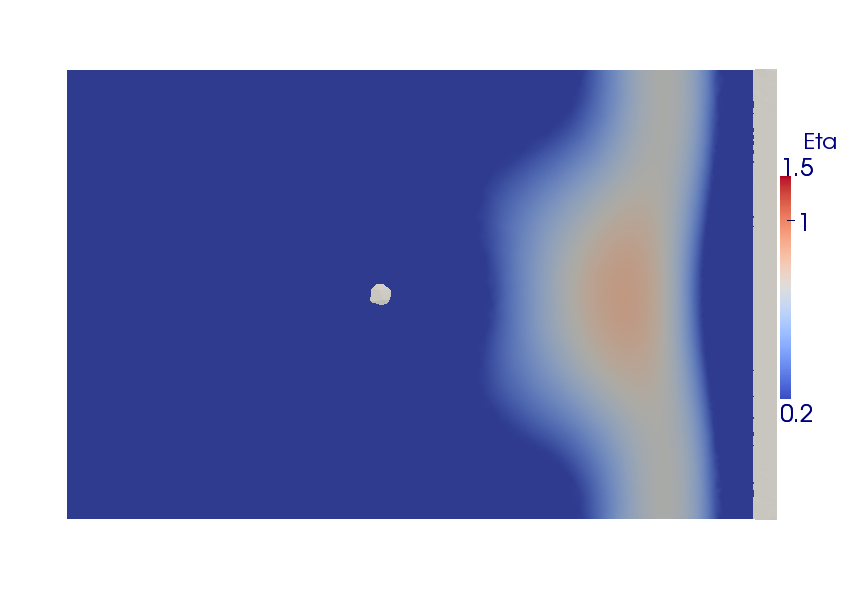}
  }\\
    \subfloat[][]{
    \includegraphics[width=0.45\columnwidth]{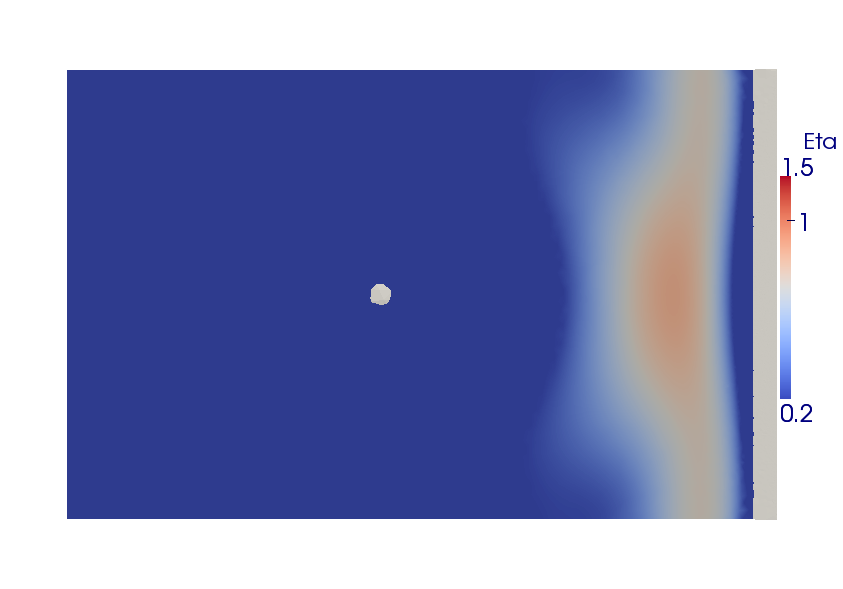}
  }
  \subfloat[][]{
    \includegraphics[width=0.45\columnwidth]{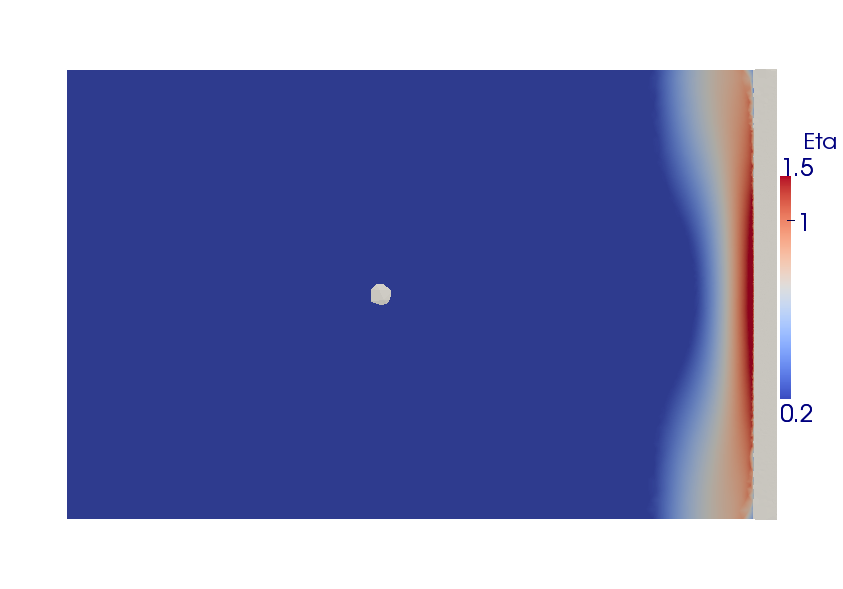}
  }
    \caption{Snapshots of the free surface elevation measured in meters as the wave passes the island and runs up the beach behind it. The island focuses the wave on its lee side and the amplified wave propagates towards the beach. The colorbar is in logarithmic scale for visualization purposes. In the present case the run-up amplification is $1.59~.$}
    \label{fig:snapshots}
\end{figure*} 

\begin{figure*}[t]
\begin{center}
  \begingroup
  \tikzset{every picture/.style={scale=.75}}
 % \centering
  \subfloat[][]{
    \includegraphics{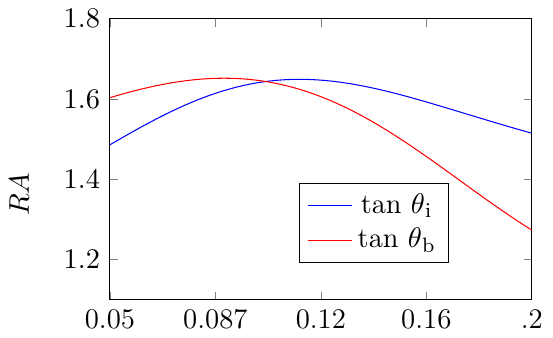}
  }
  \subfloat[][]{
    \includegraphics{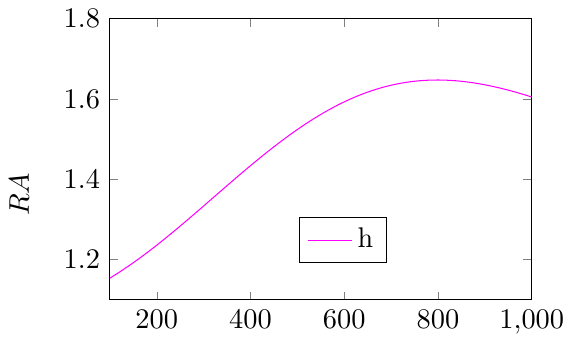}
  }\\
  \subfloat[][]{
    \includegraphics{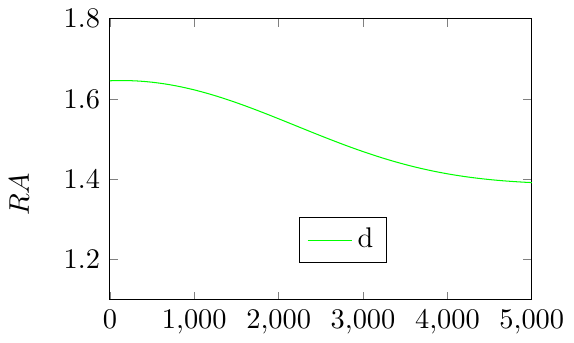}
  }
  \subfloat[][]{
    \includegraphics{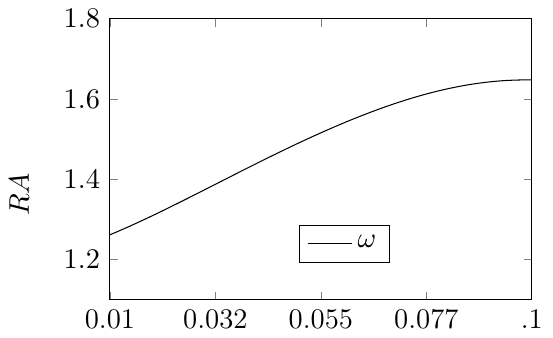}
  }
  \endgroup
  \end{center}
  \caption{Local sensitivity of the maximum run-up amplification on (a) the island and beach slopes, (b) the water depth, (c) the distance between the island and the beach and (d) the cyclic frequency of the wave. The range of the above parameters can be found in Table \ref{tab:ranges}.}
  \label{fig:sensitivity}
\end{figure*}

\begin{figure}[t]
\vspace*{2mm}
\begin{center}
\includegraphics[width=8.3cm]{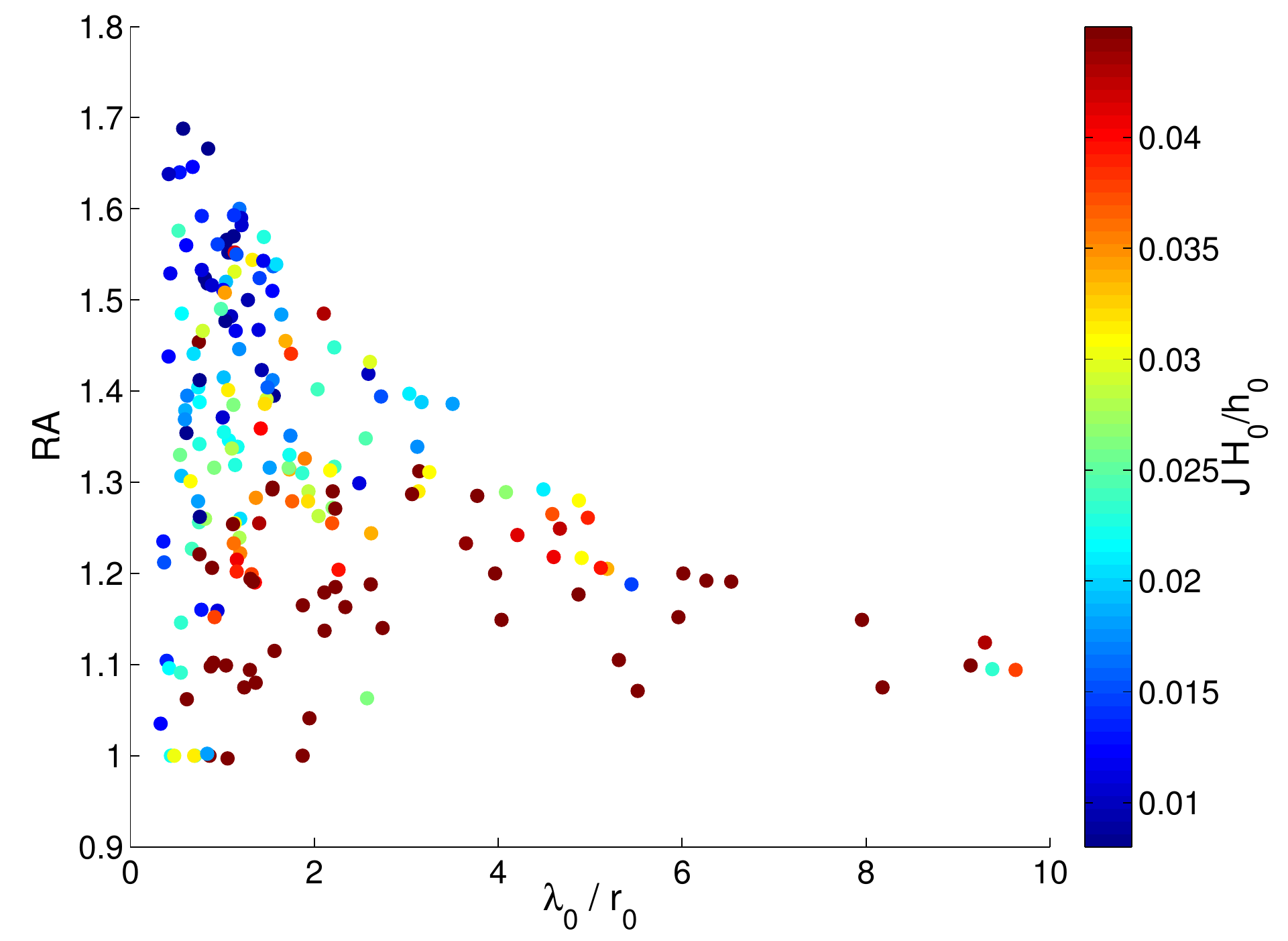}
\end{center}
\caption{Run-up amplification (RA) as a function of the wavelength to the island radius (at its base) ratio. The color code indicates the surf similarity (Iribarren number) computed with the beach slope and multiplied with the wave nonlinearity (wave height to water depth ratio). }
\label{fig:Ampli}
\end{figure}

After running $200$ simulations, we have found that in none of the situations considered the island did offer protection to the coastal area behind it. On the contrary, we have measured amplified run-up on the beach behind it compared to a lateral location on the beach, not directly affected by the presence of the island (Fig. \ref{fig:hist}). This finding shows that small islands in the vicinity of the mainland will act as amplifiers of long wave severity at the region directly behind them and not as natural barriers as it was commonly believed so far. The maximum amplification achieved was $\sim70\%$ more than were the island absent and the median amplification factor is $1.3$. The island focuses the wave on its lee side, while far from it the wave propagates unaffected (Fig. \ref{fig:snapshots}). The amplified wave propagates towards the beach and causes higher run-up in the region directly behind the island.

One of the key questions is which parameters control the run-up amplification (RA) and in what way. To answer these questions, we can use the statistical model. We perform a local sensitivity analysis around the maximum RA by fixing all parameters except one each time at the value which corresponds to the maximum RA and we vary the excluded parameter across the whole range of its input space (Fig. \ref{fig:sensitivity}). We can observe that some parameters vary more than others and thus are more important. These are the water depth, the beach slope and the cyclic frequency of the wave. Having said that, one would wonder why the parameters of the island do not seem to be that important.  The answer might not be simple, because dependencies could be hidden in the correlations of the input parameters.

To better understand these dependencies, it is of interest to recombine the input parameters in order to obtain nondimensional but physically interpretable measures. In Figure \ref{fig:Ampli} we express the RA as a function of the ratio of the wavelength over the island radius at its base  $\lambda_0/r_0$ and the Iribarren number $J$ computed using the beach slope and normalized with the relative wave amplitude $H_0/h_0$. We see that the RA  strongly depends on the ratio $\lambda_0/r_0$ and that the highest values are attained when the wavelength is almost equal to the island radius. The normalized Iribarren number gives a satisfactory classification, with smaller values qualitatively leading to higher RA. Of course the complexity of the problem is superior and cannot be completely explained by the previous two measures. Nevertheless, Figure  \ref{fig:Ampli} shows that a comprehensive knowledge of the system can give better insight than pure statistics and implies that interdisciplinary problems like this one should be treated with close collaboration between the various fields.

\section{Conclusions}  %% \conclusions[modified heading if necessary]

We examined the effect the presence of a small conical island has on the long-wave run-up on a plane beach behind it. Using a simplified geometry dependent on five physical parameters, we wanted to find the combination of parameters which will give us the maximum run-up amplification with a minimal computational cost. To achieve that, we employed an active experimental design strategy based on a Gaussian Process surrogate model. The strategy, which is parallelizable, can handle efficiently the tradeoff between exploration of the input space and focusing on the region where the $\argmax_{x} f (x)$ is believed to reside in. 

Even though our algorithm is asymptotically convergent, we are interested in its behavior for a finite time horizon $T$. Comparing our strategy to other commonly used ones, we showed that it performs better in most cases. Overall, the active experimental design approach can reduce the computational cost more than $60\%$ compared to a classic experimental design (LHS) and potentially much higher (e.g. Fig. \ref{fig:expe_gaussian}). In addition, the computational gain is orders of magnitude smaller than a regular grid approach - $3$ orders of magnitude for a $5$-dimensional problem. 

Moreover, a stopping criterion was presented, which can signal the achievement of the optimization objective and thus the end of the experiments. The development of such a criterion is essential in real applications where only a small number of experiments is allowed due to cost constraints and thus the theoretical asymptotic convergence is useless. Our stopping criterion is based on the difference in the ranking of the predictions of the surrogate model between two consecutive iterations. Even though, it is shown to correlate well with the regret $r_t^K$ (Fig. \ref{fig:regret_rank}), it depends on an empirically set threshold. Therefore, more research is needed to develop a more robust stopping criterion, which will either be derived directly from the learning algorithm or will relate the threshold to the dimensionality of the problem. 

The active learning strategy is not restricted to tsunami research and can be applied to a wide range of problems and disciplines where the optimization should be balanced with a reasonable computational or actual cost. Another interesting perspective is to incorporate in the optimization not only physical parameters, but also numerical ones, such as the spatial discretization, the placement of the (virtual) sensors and others. The inclusion of these numerical parameters can be handled by the \textsf{GP-UCB-PE} algorithm. Finally, further research is needed for the development of active learning algorithms for multi-objective optimization and pareto front tracking. 

From a physical point of view, our results show that for the given setup and range of input parameters, the island instead of protecting the beach behind it, as it was widely believed so far, it acts as a focusing lens of wave energy on its lee side. Until now, the prevailing practice in studying maximum run-up for civil defense applications has been that a plane beach provides the worst possible condition for wave amplification and thus, small offshore islands were believed to offer protection to coastal areas in their vicinity.This finding is of fundamental importance for the correct education of coastal communities and thus their preparedness in case of a tsunami. 
%TEXT

%\appendix
%\section{\\ \\ \hspace*{-7mm} HEADING}    %% Appendix A
%
%\subsection                               %% Appendix A1, A2, etc.

%\begin{acknowledgements}
%TEXT
%\end{acknowledgements}

\bibliographystyle{plain}
\bibliography{../tsbib}

\end{document}